\documentclass[aps,prd,preprint,amssymb,nofootinbib,showpacs,preprintnumbers,floatfix]{revtex4-1}

\usepackage{amsmath}
\usepackage{graphicx}
\graphicspath{{Figures/}}
\usepackage{multirow} % for using multirow and multicolumn in tables

\usepackage[mathcal]{eucal}

\usepackage{epsfig}
\usepackage{dcolumn} % Align table columns on decimal point
\usepackage{bm} % bold math
\usepackage{slashed}

\usepackage{feynmf}

\newcommand{\eqnref}[1]{Eq.~(\ref{eqn:#1})}

\newcommand{\secref}[1]{Sec.~\ref{sec:#1}}

\newcommand{\figref}[1]{Fig.~\ref{fig:#1}}

\begin{document}

\preprint{UCI-TR-2011-06}

\title{A $Z'$ Model for the CDF Dijet Anomaly}

\author{Felix Yu}
\email{felixy@uci.edu}
\affiliation{Department of Physics and Astronomy,
University of California, Irvine, CA 92697-4575, USA}

\date{April 1, 2011}

%%%%%%%%%%%%%%%%%%%%%%%%%%%%%%%%%%%%%

\begin{abstract}
We adopt a bottom-up approach to constructing a new physics model to
explain the CDF excess seen in dijets with an associated lepton and
missing transverse energy.  We find that the 145 GeV broad feature
seen by CDF in the dijet invariant mass distribution can be explained
by a $Z'$ boson with a mass of 145 GeV that couples only to first
generation quarks.  After dijet resonance constraints are considered,
a sizeable region of the parameter space favored by the CDF anomaly
remains viable.

\end{abstract}

%11.10.Kk Field theories in dimensions other than four
%11.25.Mj Compactification and four-dimensional models
%11.30.Hv Flavor symmetries
%11.30.Pb Supersymmetry
%11.30.Qc Spontaneous and radiative symmetry breaking

%12.15.Ff Quark and lepton masses and mixing
%12.15.Hh Determination of Kobayashi-Maskawa matrix elements
%12.60.Fr Extensions of electroweak Higgs sector 
%12.60.Jv Supersymmetric models

%13.85.-t Hadron-induced high- and super-high-energy interactions (> 10 GeV)
%13.85.Ni Inclusive production with identified hadrons
%13.85.Qk Inclusive production with identified leptons, photons, or other 
%         nonhadronic particles 

\maketitle

%%%%%%%%%%%%%%%%%%%%%%%%%%%%%%%%%%%%%

\section{Introduction}
\label{sec:introduction}
Recently, using 4.3 fb$^{-1}$, the CDF collaboration reported a 3.3
sigma excess over Standard Model (SM) background in dijet events with
an associated lepton (electron or muon) and missing
energy~\cite{Aaltonen:2011mk}.  This excess is present in the dijet
invariant mass range of 120--160 GeV, and a Gaussian fit to the
background-subtracted histogram in this mass range gives a Gaussian
peak at $144 \pm 5$ GeV.  While this anomaly is interesting in its own
right, a confirmation by the D0 collaboration or the persistence of
this anomaly as the Tevatron accumulates its final dataset would
provide theorists with a robust signal of new physics (NP) beyond the
SM.  The gradual accumulation of such anomalies in collider data from
the Tevatron and the LHC will serve as the bedrock for extending
physics beyond the SM.  While many of the NP models in the past few
decades have been constructed from top-down approach, in this
instance, we can adopt a bottom-up experimentally driven construction
of a NP model.

From a bottom-up approach, we look for minimal extensions to the SM in
terms of both new field content and new symmetries.  While this
approach will not in general lead to the construction of a full model,
the resulting data-driven minimal model can readily be embedded as a
feature of complete, top-down new physics constructions.  This
embedding is key to identifying future searches and cross channels
that will validate or disprove the proposed full model.

In this paper, we consider possible new physics explanations for the
CDF excess in dijet events with an electron or muon and missing
energy.  In~\secref{approach}, we discuss our bottom-up approach to
construct the simplest possible NP model that can give rise to the
observed excess, which we find to be a $Z'$ model with a $Z'$ mass of
about 150 GeV.  In~\secref{bounds}, we briefly discuss the collider
constraints on $Z'$ masses and couplings and conclude that
flavor-universal $Z'$ models that could explain the CDF anomaly are
excluded.  We therefore discard flavor-universality and consider the
$Z'_{ud}$ model, a $Z'$ that couples equally to and only to the first
generation quarks, which we present in~\secref{model}.  We conclude
in~\secref{conclusion} with a summary and a brief discussion of
possible cross-channels to check at the Tevatron or the LHC.

\section{Approach}
\label{sec:approach}
New physics parameter space is large, and many different models can be
made to fit the excess.  In our bottom-up approach, we seek minimal
extensions of the SM, keeping in mind both theoretical and
experimental constraints on such extensions.  

The CDF event selection calls for events with one electron (muon) with
$E_T (p_T) > 20$ GeV and no other leptons with $p_T > 10$ GeV; in
addition, a $Z$ mass window (from $76--106$ GeV) cut on dilepton
candidate events is imposed.  The event selection also requires
strictly two jets, reconstructed using a fixed-cone algorithm with
$\Delta R = \sqrt{ (\Delta \eta)^2 + (\Delta \phi)^2} = 0.4$, each
with $E_T > 30$ GeV and $|\eta| < 2.4$: the dijet system must have
$p_T \geq 40$ GeV.  In addition, events are required to have missing
transverse energy (MET) $\slashed{E}_T > 25$ GeV.  The transverse mass
of the single hard lepton and the MET is required to be compatible
with a $W$-candidate, $m_T^W = \sqrt{2 p_T^\ell \slashed{E}_T (1 -
  \cos (\Delta \phi_{\ell \nu}))} \geq 30$ GeV.  Additional details
regarding jet energy corrections and isolation requirements are given
in~\cite{Aaltonen:2011mk} (cf.~Table 4.2 and Section 8.1 of the
Cavaliere thesis).

The event excess is present in both electrons and muons.  For
electrons, the excess number of events is $156 \pm 42$, and for muons,
the excess is $97 \pm 38$.  Naively summing the systematic errors in
quadrature, we find the total excess is $256 \pm 56.6$ events.  The
new physics contribution thus needs to give an excess in the dijet,
single hard lepton, and MET final state with an effective cross
section (new physics cross section $\times$ acceptance) of about 60 fb
for the Tevatron collider.  If we presume the lepton arises from a $W$
boson, the required effective cross section for NP production
including a hard $W$ boson is about 270 fb, and if we estimate the
acceptance factor for our signal to be about 10\%, then we are looking
for a total signal cross section of about 2.7 pb.  For comparison, the
measured $WW/WZ$ cross section is 18.1 $\pm$ 3.3 (stat.) $\pm$ 2.5
(syst.) pb, consistent with the SM prediction of 15.9 $\pm$ 0.9
pb~\cite{Aaltonen:2011mk}.  Thus, we aim to develop a minimal new
physics model that has approximately an $\mathcal{O} (1 \text{ pb})$
production cross section, including $W$ emission.

There are two main issues from a bottom-up perspective.  First,
considering the excess in the dijet invariant mass distribution from
120--160 GeV, we can interpret it as a colored resonance, an uncolored
resonance, or a kinematic feature from a cascade decay.  Second,
concerning the presence of the hard lepton, we can consider, in turn,
scenarios that are lepton number violating, lepton number conserving
but flavor violating, or lepton number and flavor conserving with the
separate possibility of kinematic suppression of additional leptons.
Separately, the observed MET could arise from SM neutrinos, or it
could arise from a NP source of missing energy: in the first case, we
could again consider possible NP scenarios of lepton flavor and/or
number violation, but this is redundant and unnecessary given the hard
lepton.  We will first discuss the dijet invariant mass excess as a
possible kinematic feature from a cascade decay chain.

\subsection{Cascade decay chain explanation for the $m_{jj}$ excess}
Invariant mass distributions from cascade decay chains can appear to
have broad resonance features when the underlying particle masses are
tuned appropriately and the correct particle combinations are
isolated~\cite{Allanach:2000kt,Rajaraman:2010hy}.  A simple example of
such a cascade decay chain is when a massive color octet decays via a
on- or off-shell massive color triplet to a color singlet that
subsequently escapes the detector: in supersymmetry (SUSY), this is
the familiar gluino cascade decay, $\tilde{g} \rightarrow q \tilde{q}
\rightarrow q q \tilde{\chi}_1^0$, which can have a large rate if the
$m_{\tilde{\chi}_1^0} < m_{\tilde{g}}$.  For example, if
$m_{\tilde{g}} = 420$ GeV, $m_{\tilde{q}} = 380$ GeV, and
$m_{\tilde{\chi}} = 150$ GeV, the exact dijet invariant mass edge
would be 164 GeV and the dijet invariant mass distribution would
exhibit the usual triangular shape, assuming the emitted quarks are
massless.  Since the quarks shower and hadronize, however, we expect
the triangular feature to be smoothed out and the distribution to have
a tail from jet-parton momenta mismatch as well as pollution by wrong
dijet combinations.

There are several difficulties with making such a possibility work.
First, generating a lepton together with this dijet invariant mass
feature requires additional ingredients.  If we assume the lepton
arose from the same decay chain, we can consider an illustrative SUSY
example:
\begin{equation}
\tilde{g} \rightarrow q \tilde{q} \rightarrow q q \tilde{\chi}_1^\pm
\rightarrow q q \ell^\pm \tilde{\nu} \ .
\label{eqn:gludecay}
\end{equation}
In this SUSY example, we would need to have large $R$-parity violation
in order to singly produce the $\tilde{g}$, but we would also need to
minimize $R$-parity violation in order to force the prescribed decay
chain.  If we retain $R$-parity and still assume the $\tilde{\nu}$ is
the lightest supersymmetric particle (LSP), we can assume the gluino
is either produced in pairs or in association with a squark.  In
either case, the searches for SUSY in final states of one lepton,
jets, and MET~\cite{Aad:2011hh} or opposite-sign dilepton events,
jets, and MET~\cite{Chatrchyan:2011bz} have put strict constraints on
gluinos and squarks with masses below about 500 GeV (and even up to
700 GeV).  

If we make $\tilde{\chi}_1^0$ the LSP, then the lepton could minimally
arise from the leptonic decay of a $W$ boson: such a $W$ could be
produced from a heavy squark to light squark decay or in a chargino to
neutralino decay (or vice-versa).  An example SUSY process for the CDF
excess in this case is
\begin{equation}
\tilde{q} \tilde{g} \rightarrow (q_1 \tilde{\chi}_1^\pm) (q_2 \tilde{q})
\rightarrow (q_1 W \tilde{\chi}_1^0) (q_2 q_3 \tilde{\chi}_1^0) \ ,
\label{eqn:sqgludecay}
\end{equation}
where the $W \rightarrow \ell \nu$, $q_1$ is soft, and $q_2$ and $q_3$
are hard jets that give the invariant mass excess.  Here, in contrast
with the above gluino decay in~\eqnref{gludecay}, the necessary
addition of a weak gauge coupling is a model-independent penalty in
order to incorporate the $W$ in a cascade decay chain, and the
$\approx 22$\% leptonic branching ratio of the $W$ boson is an
additional penalty.  An alternative to~\eqnref{sqgludecay} is if a
slepton were part of the cascade process, but such decay chains
typically give rise to large multilepton signals, which are disfavored
from~\cite{Aad:2011hh} and~\cite{Chatrchyan:2011bz}.  Moreover,
cascade processes such as~\eqnref{sqgludecay} can be easily checked in
the jets+MET cross-channel, and recent results~\cite{Alves:2010za,
  Khachatryan:2011tk, daCosta:2011qk} on these final states indicate
that such spectra would need fine-tuning in order to evade
constraints.  Other choices for the stable LSP besides a neutralino or
sneutrino are ruled out or disfavored from the recent searches for
long-lived massive charged particles~\cite{Abazov:2008qu,
  Aaltonen:2009kea, Aad:2011mb}.

To summarize, a SUSY decay chain explanation for the dijet excess
suffers from two competing considerations.  In order to have an
appreciable SUSY colored cross section at the Tevatron, we must make
the gluinos and squarks relatively light.  Yet, the requirement to
have a lepton emitted in a cascade requires a slepton, a sneutrino, or
a $W$ insertion, making the resulting effective cross section for a 2
jets + lepton + MET final state disfavored given ATLAS and CMS
searches.  Since there is a great deal of freedom in arranging the
SUSY spectrum, however, we do not rule out a SUSY explanation for the
CDF anomaly but instead leave such a construction for future work.

We can also consider a non-SUSY decay chain explanation.  The simplest
would be a non-SUSY version of~\eqnref{gludecay}, which minimally
requires the introduction of a new heavy color octet $X$ and
triplet $Q$,
\begin{equation}
X \rightarrow q Q \rightarrow q q W \rightarrow q q \ell \nu \ .
\label{eqn:nonSUSYdecay}
\end{equation}
We note that other color representations for the initial particle are
also obviously possible: the only assumption is that the $W$ at the
end of the decay arises from a weak coupling vertex involving a SM
quark and some new physics particle, which must necessarily be in the
triplet representation of $SU(3)$.  For example, the new $Q$ can be
considered as a fourth generation quark, though no such assumption is
truly motivated from the bottom-up approach.  We note that although
this decay chain readily produces all of the final state particles of
the CDF anomaly, the decay chain arises from a resonance, and so must
couple directly to quarks and/or gluons.  This implies the constraints
and phenomenology are similar to the dijet resonance considerations,
and so we will incorporate this discussion with the next subsection.

\subsection{Resonance decaying to two jets}
A more straightforward bottom-up construction is to hypothesize the
two jets arise from a resonance, not a cascade decay chain.  Given the
lack of $b$-tagging information about the jets, we note the resonance,
if colored, could be one of many different color representations under
$SU(3)$.  In addition, the emission of the hard lepton and the MET
requirement allows one of several possibilities: NP could be lepton
number violating (LNV), lepton flavor violationg (LFV), or lepton
number and flavor conserving.  Because the MET requirement is small,
we can naturally associate the MET to be a neutrino emission and
confine our discussion to NP scenarios that conserve lepton number and
lepton flavor.  We leave the possible construction of a viable LNV or
LFV model for future work.  Therefore, we consider a new physics
resonance that decays to two jets where the decay chain includes a $W$
boson, or the resonance is produced in association with a $W$ boson.
The Tevatron production cross section then necessarily includes weak
coupling and the $W$ leptonic branching fraction penalty: to avoid
this, we could instead consider a $W'$ boson that decays favorably to
leptons.  Recent constraints on a new $W'$ boson with leptonic
couplings, however, completely exclude any such $W'$ boson with a mass
below about 1.5 TeV~\cite{:2007bs, Aaltonen:2010jj,
  Khachatryan:2010fa, Chatrchyan:2011dx, Aad:2011fe}.

The main difficulty with the resonance + SM $W$ model is that, by
construction, the new resonance can always be produced in an
$s$-channel process without an associated $W$ boson and hence is
subject to direct searches for dijet resonances.  Correspondingly, the
dijet resonance cannot be strongly coupled (the dijet search
constraints are discussed in~\secref{bounds}), but even so, we are
left with many possibilities for the couplings and character of the
new resonance.  The resonance can be colored or uncolored, can couple
exclusively to gluons, quarks, or both, and can conserve or violate
quark flavor.  We will not consider a resonance coupling exclusively
to gluons, because we require the resonance to be produced in
association with a $W$ boson.  Similarly, we will not consider a
fractionally-charged resonance with a gluon-(anti-)quark coupling
because the coupling would be non-diagonal if the resonance were in
the same $SU(3)$ representation as the (anti-) quark, and for higher
$SU(3)$ representations, the gluon-quark resonance would require a
careful consideration of constraints to ensure it remains viable.  We
reserve a study of phenomenology and constraints of this interesting
quark--gluon resonance model for future work.

If the resonance has quark-(anti-)quark couplings, we could expect the
dominant process for associated $W$ boson production to come from the
2--2 $t$-channel scattering process
\begin{equation}
q q \rightarrow W X_{qq} \rightarrow (\ell \nu) (q q) \ ,
\label{eqn:Xqq}
\end{equation}
where the $t$-channel exchanged $SU(3)$ fundamental could also be a
fourth generation quark.  We see that this process is reminiscent
of~\eqnref{nonSUSYdecay}: in fact these production processes can be
considered as differing cases of the same underlying new physics model
that introduces a new $SU(3)$ octet $X_{qq}$ and a new $SU(3)$ triplet
$Q$.  We remark that if the $t$-channel exchange particle is a SM
quark, then the only two free parameters are the resonance-quark-quark
coupling and the width of the resonance, since the mass of the
resonance is fixed from the Gaussian fit to the dijet excess performed
by CDF.  On one hand, these two free parameters are constrained by
direct dijet searches, and on the other hand, these are the only
parameters available to ensure the cross section for resonance + $W$
production matches the observed number of events at CDF.  If the
$t$-channel also included a fourth generation quark, however, then we
have additional freedom to modify separately the direct dijet cross
section and the resonance + $W$ production cross section.

Alternatively, for the resonance with quark-(anti-)quark couplings,
the dominant process for $W$ emission could be from $s$-channel
production, as in
\begin{equation}
q q \rightarrow W' \rightarrow W Z' \rightarrow (\ell \nu) (q q) \ .
\label{eqn:Wprime}
\end{equation}
Here, we have changed the resonance notation to the traditional $Z'$
to emphasize that it is a color singlet.  Again, the advantage of this
$s$-channel construction is there are separate couplings that control
the CDF event excess and the direct dijet production of $Z'$: this
model freedom can clearly be used to evade constraints from direct
searches while also ensuring the correct production cross section for
the excess.  A similar process to~\eqnref{Wprime} would be production
of a techni-rho decaying to a techni-pion: $q q \rightarrow \rho_{TC}
\rightarrow W \pi_{TC} \rightarrow (\ell \nu) (q q)$.

The $s$-channel explanation is disfavored, however, from both the CDF
data (cf.~Fig.~9.13 of \cite{Aaltonen:2011mk}) and the bounds on new
dijet resonances (discussed in detail in~\secref{bounds}).  In the CDF
analysis, they do not find a resonance feature in the $jj \ell \nu$
invariant mass; instead, the total invariant mass is consistent with
the background hypothesis.  We note that we can avoid generating a
feature in $m_{jj \ell \nu}$ by postulating additional decay products
that are invisible, soft, or otherwise missed by the detector.  Such
constructions and their corresponding experimental constraints are
very model dependent, however, so following our bottom-up approach, we
consider $t$-channel production of a $Z'$ resonance + SM $W$ boson
with only SM quarks exchanged.  We will find that we can successfully
fit the CDF dijet excess with such a $Z'$ model.

\begin{figure}[tbhp]
\includegraphics[scale=0.60]{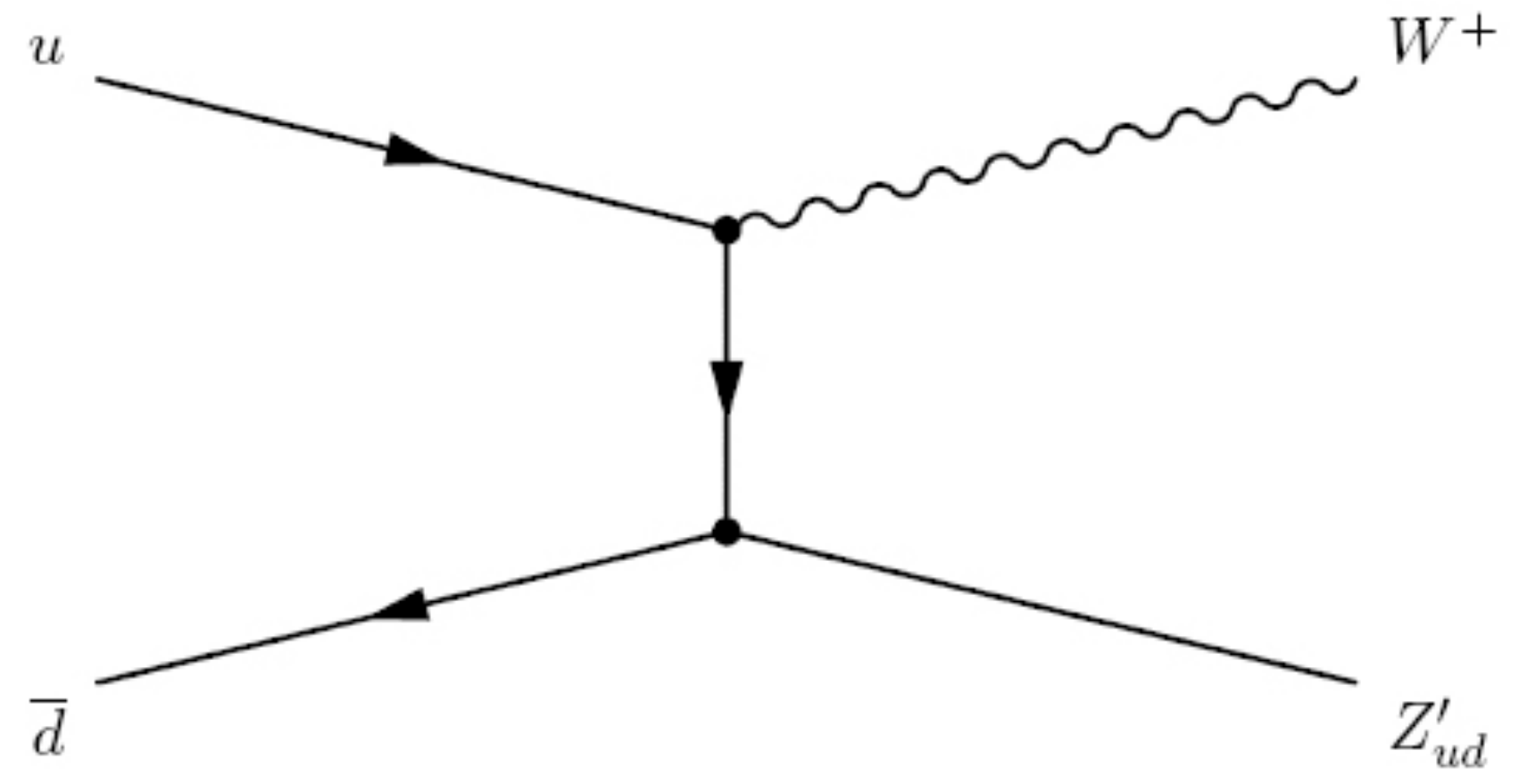}
\caption{\label{fig:tchannel} $W^+ Z'_{ud}$ associated production with
  a $t$-channel SM $d$ quark.}
\end{figure}

In summary, from a bottom-up perspective, we discussed the possibility
of a dijet cascade decay invariant mass feature and a dijet resonance.
We also considered the origin of the observed lepton: if the lepton
comes from a $W$ boson, then the full cross section would require a
weak coupling and pay a price in the $W$ leptonic branching ratio.  On
the other hand, if the lepton is from a cascade decay, then
model-dependent tuning is needed to ensure a large branching fraction
for a single lepton.  Given these considerations, we find the simplest
new physics model for explaining the CDF anomaly is a $Z'$ dijet
resonance produced in association with a $W$ that decays leptonically,
as shown in~\figref{tchannel}.

\section{Bounds on a new $Z'$ boson}
\label{sec:bounds}
From our bottom-up approach, the simplest model to explain the CDF
anomaly is a $Z'$ dijet resonance produced in association with a
leptonically decaying $W$ boson.  Constraints on such a light $Z'$,
however, are very stringent.

For instance, a $Z'$ boson with SM $Z$ couplings to leptons is ruled
out below a mass of 1071 GeV~\cite{Aaltonen:2011gp}.  From a bottom-up
approach, however, we do not require the $Z'$ to couple to leptons,
and hence we can postulate that the $Z'$ is leptophobic,
{\textit{i.e.}} only couples to quarks.  Even so, searches for dijet
resonances place strong constraints on this type of $Z'$ boson.  The
most recent results from ATLAS~\cite{:2010bc} and
CMS~\cite{Khachatryan:2010jd}, however, look for dijet invariant
masses above $200$ GeV and $220$ GeV, respectively, in order to avoid,
presumably, the QCD background contamination in the low dijet mass
region.  These hard cuts will clearly discard our light $Z'$ events.
Similarly, CDF and D0 searches applicable to leptophobic $Z'$ bosons
have dijet mass thresholds of at least 180 GeV~\cite{Abe:1995jz,
  Abe:1997hm, Abazov:2003tj, Melnitchouk:2008ij, Aaltonen:2008dn} or
look for $t \overline{t}$ resonances~\cite{:2007dz}.  All of these
constraints apply to $Z'$ masses outside our range of interest, given
the Gaussian fit of CDF's dijet excess has a mean 144 GeV.

We find the relevant experimental constraint on dijet resonances in
this mass range comes from the UA2 collaboration~\cite{Alitti:1993pn}.
In this analysis, they assumed a $Z'$ with exactly SM couplings and a
width that scaled with the mass ratio of the $Z'$ to the $Z$.  In
addition, the cross section was also corrected with a K-factor of
about 1.30~\cite{Hamberg:1990np}, based on the K-factor calculated for
SM Drell-Yan $Z$ production.  Their results concluded that a SM-like
$Z'$ is excluded at 150 GeV.

Although our naive leptophobic $Z'$ model with SM $Z$ couplings to
quarks is ruled out from UA2 data, we can choose to abandon flavor
universality.  In this case, we expect the UA2 bound implies our $Z'$
coupling needs to be about $1/ \sqrt{2}$ weaker than the SM $Z$
coupling to quarks.  If we retain flavor universality, we can satisfy
the UA2 constraint and produce the desired CDF excess with a
$g_{\text{universal}} = 0.20-0.25$, but we would also need to
consider constraints from Higgs searches in the $\ell \nu b b$ final
state~\cite{Aaltonen:2009pj}.  We will therefore consider a $Z'$ that
only couples to up and down quarks with equal couplings, avoiding
flavor constraints.  This is our minimal model motivated from a
bottom-up approach to the CDF dijet, lepton, and MET events excess.
Taking into account the $Z'$ mass will be fixed by the CDF dijet
Gaussian fit, this model has two free parameters: $g_{ud}$ and the
$Z'_{ud}$ width.

\section{The $Z'_{ud}$ model and simulation}
\label{sec:model}
Faced with the severe constraint from UA2 on light $Z'$ bosons in the
150 GeV mass range, we construct the $Z'_{ud}$ model with the
Lagrangian
\begin{equation}
\mathcal{L} \supset - g_{ud} Z'_{ud {\mu}} \overline{u} \gamma^{\mu} u
- g_{ud} Z'_{ud {\mu}} \overline{d} \gamma^{\mu} d \ ,
\label{eqn:Lagrangian}
\end{equation}
where $Z'_{ud}$ is a new $U(1)'$ gauge boson, and $g_{ud}$ is the new
coupling, same for both up and down quarks.  The $Z'_{ud}$ is a
singlet under the Standard Model gauge group and we turn off its
mixing with the SM $Z$ boson.  (For a recent review on $Z'$ models and
phenomenology, see~\cite{Langacker:2008yv} by P.~Langacker.)

For our purposes, we can consider the $Z'_{ud}$ as a particular
leptophobic $Z'$ model based on gauging the SM baryon number symmetry.
While additional field content is needed to cancel anomalies, such a
full model description would follow the earlier work along the lines
of~\cite{Foot:1989ts, Carena:2004xs, FileviezPerez:2010gw}.

\begin{figure}[tbhp]
\includegraphics[scale=0.60]{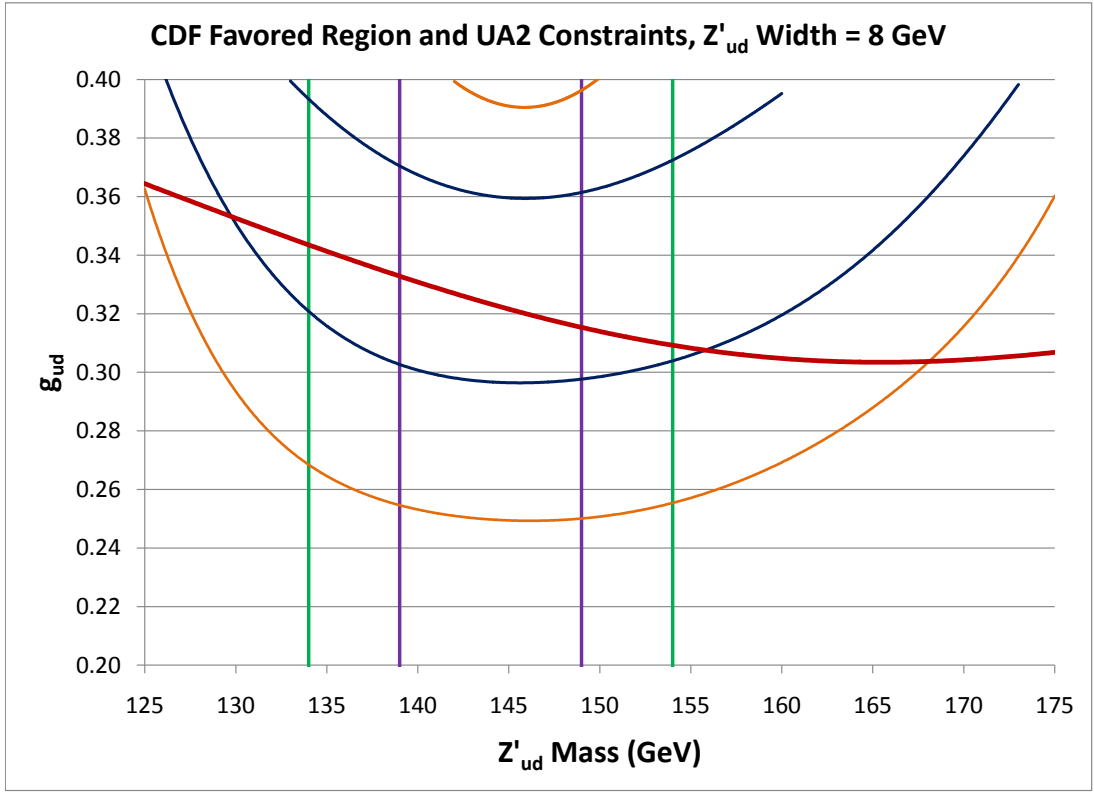}
\caption{\label{fig:Zpud-8} (color online).  The blue (orange) curves
  are the 1 (2) $\sigma$ bounds on $g_{ud}$ coupling for given $Z'$
  mass and obtained from matching the observed number of excess events
  seen at CDF (see text).  The purple (green) vertical lines indicate
  the 1 (2) $\sigma$ limits of the $Z'$ mass from the Gaussian fit of
  $m_{jj}$ performed by CDF.  The red curve indicates the extracted
  limit (at LO) on the coupling $g_{ud}$ from the UA2 search for
  SM-like $Z'$s decaying to two jets~\cite{Alitti:1993pn} (see text).
  The $Z'$ width is fixed to be $8$ GeV for entire mass range.}
\end{figure}

\begin{figure}[tbhp]
\includegraphics[scale=0.60]{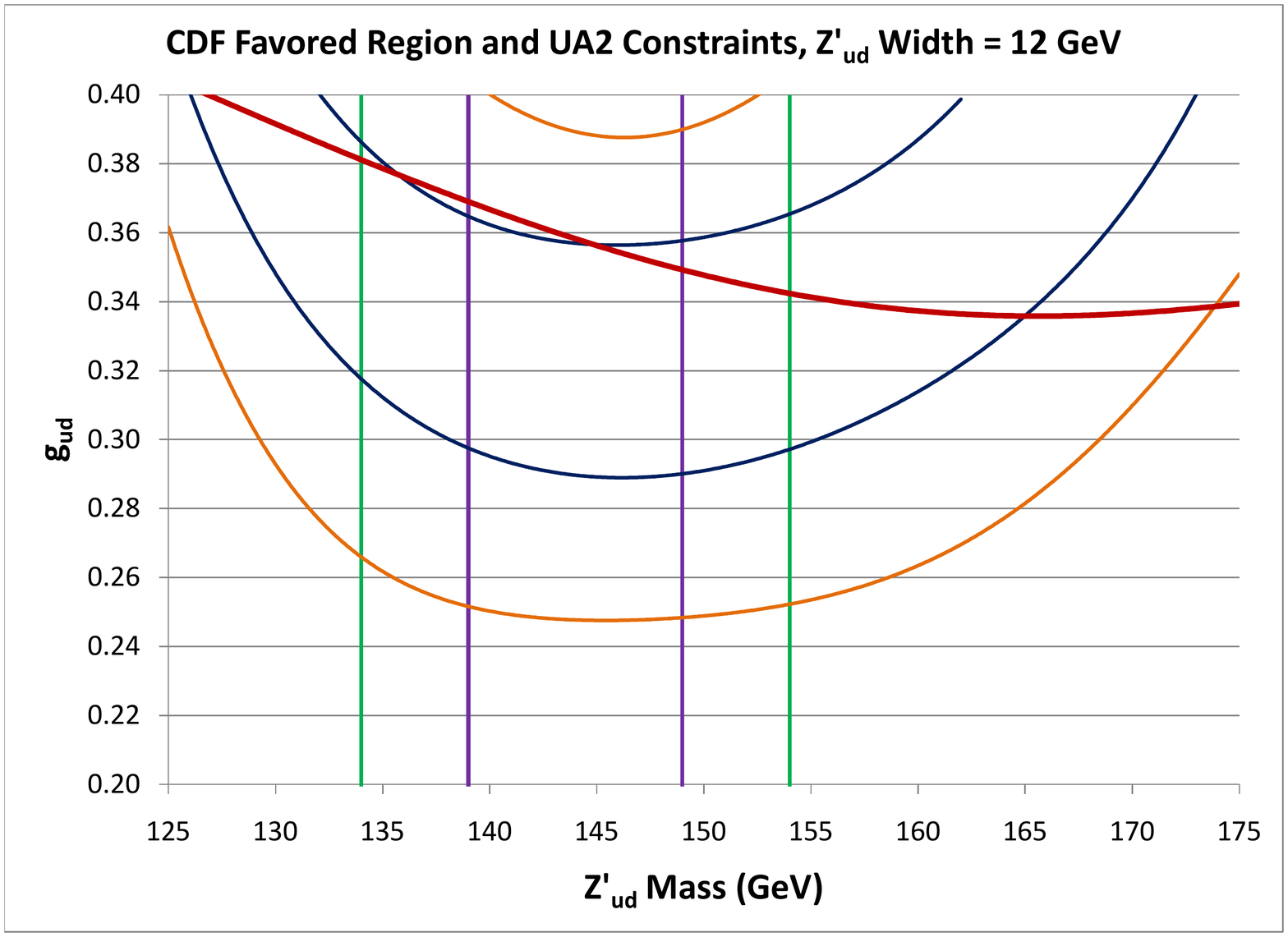}
\caption{\label{fig:Zpud-12} (color online).  Same as~\figref{Zpud-8},
  except for a $Z'$ with a width of $12$ GeV.}
\end{figure}

We simulate the $Z'_{ud}$ production for masses between 125 GeV to 175
GeV, couplings $g_{ud}$ at leading order (LO) from 0.20 to 0.40, and a
$Z'_{ud}$ width of 8 GeV or 12 GeV, using MadGraph 5
v.0.5.1~\cite{mg5} and MadEvent v.4.4.56~\cite{Alwall:2007st,
  Alwall:2008pm, Maltoni:2002qb} interfaced with Pythia
6.4.20~\cite{Sjostrand:2006za} and PGS 4~\cite{pgs}.  For testing the
match to the CDF excess, we generate Tevatron $p \overline{p}$
collisions at $\sqrt{s} = 1.96$ TeV, $p \overline{p} \rightarrow W^\pm
Z'_{ud}$, use the Pythia interface to decay, shower, and hadronize,
and then perform rough clustering and detector simulation using PGS.
We apply identical cuts as the CDF analysis~\cite{Aaltonen:2011mk}.
At each point of mass, coupling, and width, we count the number of
events within the signal region of $120$ GeV $ < m_{jj} < 160$ GeV.
Based on this event count and the Gaussian fit performed by CDF, we
find the best fit point is at about a $Z'$ mass of 144 GeV and a
coupling $g_{ud} \sim 0.33$, irrespective of the $Z'$
width, see~\figref{Zpud-8} and~\figref{Zpud-12}.

We also need to calculate the UA2 constraint~\cite{Alitti:1993pn} for
this $g_{ud}$ coupling v. $Z'_{ud}$ mass plane.  To do so, we simulate
each model point for S$p\overline{p}$S collisions of $p \overline{p}$
at $\sqrt{s} = 630$ GeV to get a (LO) $Z'_{ud}$ $s$-channel production
cross section estimate.  We also calculate the $Z'$ cross section
limit from Fig.~5 of~\cite{Alitti:1993pn}.  Based on the $g_{ud}^4$
scaling of the cross section, we can get a (LO) constraint on the
$g_{ud}$ coupling allowed by the UA2 search.  Our results are
displayed in~\figref{Zpud-8} for a $Z'_{ud}$ with an 8 GeV width
and~\figref{Zpud-12} for a 12 GeV width.  We used BRIDGE
v.2.21~\cite{Meade:2007js} to calculate the $Z'$ width for each point
to ensure the partial width of $Z' \rightarrow u \overline{u}, d
\overline{d}$ stayed below 8 GeV.  As a point of comparison, for a
$Z'$ mass of 145 GeV, $g_{ud} = 0.35$, the calculated $Z'$ width to
quarks is $2.824$ GeV.  Additional invisible decay modes would need to
added in a full model to account for the remaining $Z'$ width.

Our results demonstrate that the CDF anomaly can be favorably fit with
a $Z'_{ud}$ of mass between about 140 GeV and 150 GeV and a coupling
of $0.30 \lesssim g_{ud} \lesssim 0.36$.  For a $Z'_{ud}$ width of 8
GeV, however, slightly more than half of this favored region is
excluded by UA2.  If we increase the $Z'_{ud}$ width to 12 GeV,
though, the UA2 constraint eliminates only a small part of this
favored region.

We note that we did not include K-factors for the Tevatron and UA2
production cross sections used in~\figref{Zpud-8}
and~\figref{Zpud-12}, while the UA2 collaboration did include NLO
K-factors in their SM-like $Z'$ exclusion limit contour.  Clearly, a
full calculation of the NLO K-factors for $Z'_{ud}$ $s$-channel and
$W^\pm Z'_{ud}$ associated production is beyond the scope of this
work.  We naively expect, however, the NLO K-factor for $s$-channel
$Z'_{ud}$ production to be about 1.30, given the work
of~\cite{Hamberg:1990np}, which should rescale the UA2 exclusion curve
down by about 6.5\%.  In this case, even if no K-factor enhancement to
$W^\pm Z'_{ud}$ production at Tevatron is assumed, our conclusions
remain the same and much of our favored region is left intact.

\section{Conclusions and future searches}
\label{sec:conclusion}

We have performed a bottom-up analysis of the excess events in the
dijet, lepton, MET final state seen by CDF.  After discussing possible
new physics constructions that could explain the excess, we found a
minimal model that satisfied all present collider constraints and had
a minimal number of free parameters.  The $Z'_{ud}$ model introduces a
new $Z'$ gauge boson that only couples to first generation quarks.  We
calculated the exclusion curves for this model for two different
$Z'_{ud}$ widths and found that a significant portion of the CDF
favored region was not excluded from the UA2 dijet constraint.

It remains to identify possible cross-channels for checking the
validity of this model.  While the CDF author acknowledges the entire
anomaly may be an underestimated background (see Chapter 9 of the
Cavaliere thesis~\cite{Aaltonen:2011mk}), a search in the same
exclusive dijets, lepton, MET channel by D0 would certainly
corroborate or refute the excess.  A dijet signal from Drell-Yan
production of the $Z'_{ud}$ boson may be lost in the QCD background at
Tevatron and the LHC, but a signal may be recoverable if the
backgrounds are very well understood.  Separately, because the lepton
here arises from associated $W$ production, a smaller penalty in cross
section would be achieved by looking for a photon + dijet
signal~\footnote{We thank Tao Han for bringing up this point.}.  At
the LHC, since exclusive dijet searches are expected to be difficult
because of the QCD background, one possible search channel is in the
exclusive four jets final state.  Using MadEvent 4.4.56, we estimate
the LO cross section for di-$Z'_{ud}$ production at the 7 TeV LHC with
$g_{ud} = 0.30$ and a width of 12 GeV is 1.51 pb, which could be
testable with an integrated luminosity of $\mathcal{O}(1 \text{
  fb}^{-1})$.  Since we have an estimate for the $Z'_{ud}$ mass, the
QCD background can readily be subtracted out from a sideband
subtraction method, and wrong combinatorics can be removed from a mass
window cut and a dijet $p_T$ requirement~\cite{Aaltonen:2011mk,
  Rajaraman:2010hy}.  Additional search channels may also be
available, but their presence would be motivated from the particular
full model completion of the $Z'_{ud}$ minimal model presented here.
In a future work, we will consider possible full model completions and
differentiate their phenomenology.

Recent work that also discussed light dijet resonances
include~\cite{Bai:2010dj, Chanowitz:2011ew, Buckley:2011vc}.  In
particular, we note a very similar model was considered
in~\cite{Buckley:2011vc}.

\section*{Acknowledgements}
\label{sec:acknowledge}

FY would like to thank Arvind Rajaraman, Tim Tait, Shufang Su, Vikram
Rentala, Will Shepherd, and Robert Foot for several useful
discussions, and especially A.~Rajaraman and T.~Tait for comments on
the draft.  FY is supported by a 2010 LHC Theory Initiative Graduate
Fellowship, NSF Grant No. PHY-0705682.  FY is also supported by NSF
Grant No. PHY-0653656 and PHY-0970173.

%%%%%%%%%%%%%%%%%%%%%%%%%%%%%%%%%%%%%

\end{document}